\documentclass[aps,prb,twocolumn,superscriptaddress]{revtex4}
\usepackage{graphicx}
\usepackage{calc}
\usepackage{bm}

\begin{document}

\title{Observation of two relaxation mechanisms in transport
between spin split edge states at high imbalance}

\author{E.V.~Deviatov}
\email[]{dev@issp.ac.ru}
 \affiliation{Institute of Solid State
Physics, Chernogolovka, Moscow District 142432, Russia}

\author{A.~W\"urtz}
\affiliation{Laboratorium f\"ur Festk\"orperphysik, Universit\"at
Duisburg-Essen, Lotharstr. 1, D-47048 Duisburg, Germany}

\author{A.~Lorke}
\affiliation{Laboratorium f\"ur Festk\"orperphysik, Universit\"at
Duisburg-Essen, Lotharstr. 1, D-47048 Duisburg, Germany}

\author{M.Yu.~Melnikov}
\affiliation{Institute of Solid State Physics, Chernogolovka,
Moscow District 142432, Russia}

\author{V.T.~Dolgopolov}
\affiliation{Institute of Solid State Physics, Chernogolovka,
Moscow District 142432, Russia}

\author{D.~Reuter}
\affiliation{Lehrstuhl f\"ur Angewandte Festk\"orperphysik,
Ruhr-Universit\"at Bochum, Universit\"atsstrasse 150, D-44780
Bochum, Germany}

\author{A.D.~Wieck}
\affiliation{Lehrstuhl f\"ur Angewandte Festk\"orperphysik,
Ruhr-Universit\"at Bochum, Universit\"atsstrasse 150, D-44780
Bochum, Germany}

\date{\today}

\begin{abstract}
Using a quasi-Corbino geometry to directly study electron
transport between spin-split edge states, we find a pronounced
hysteresis in the $I-V$ curves, originating from slow relaxation
processes. We attribute this long-time relaxation to the formation
of a dynamic nuclear polarization near the sample edge. The
determined characteristic relaxation times are 25 s and 200 s
which points to the presence of two different relaxation
mechanisms. The two time constants are ascribed to the formation
of a local nuclear polarization due to flip-flop processes and the
diffusion of nuclear spins.
\end{abstract}


\maketitle

\section{introduction}

In a quantizing magnetic field, energy levels in a two dimensional
electron gas (2DEG) bend up near the edges of the sample, forming
edge states (ES) at the lines of intersection with the Fermi
level. According to the picture of B\"uttiker~\cite{buttiker},
transport in two dimensional electron systems takes place mostly
in ES because of a zero bulk dissipative conductivity in the
quantum Hall effect regime. This single-particle picture was
modified by Shklovskii et al.~\cite{shklovsky}, taking into
account electrostatic interactions between electrons. This leads
to the appearance of a set of incompressible and compressible
strips near the 2DEG edge. This ES picture is nowadays widely
accepted and in good agreement with experimental
results~\cite{haug}.

Several experiments were performed, investigating not only
transport along the 2DEG edge  but also inter-edge-state charge
transfer~\cite{haug}. In  charge transfer between the two
spin-split ES of the lowest Landau level, the necessity for
spin-flips diminishes the tunneling probability. For this reason,
the equilibration length between spin-split edge states can be as
high as 1 mm at low temperatures, despite the large spatial
overlap of electron wave functions~\cite{gusev,muller}.

Formerly, the electron spin-flip was attributed to spin-orbit
coupling~\cite{muller,khaetskii} while today it is known that the
electron spin-flip can be accompanied by the spin-flop of a
nucleus (so-called dynamic nuclear polarization (DNP)) near the
edge~\cite{dixon,komiyama,komiyama2}. A key feature of this effect
is a pronounced hysteresis of the $I-V$ traces due to the high
nuclear-spin-lattice relaxation time $T_1$, which was also
reported  for the bulk~\cite{dobers,berg,kane,portal}.

Most experiments on the charge transfer between ES were performed
using the so-called cross-gate technique \cite{haug1} where
current is injected into one outer ES. The resistance of this
outer ES reflects the current redistribution among all
participating ES. To see a significant effect, the size of the
interaction region has to be comparable with the equilibration
length~\cite{muller}. For this reason the cross-gate method is not
suitable to study  electron transport at high imbalance between ES
with a negligible current between them. On the other hand, new
physical effects such as a hysteresis due to the switching of the
positions of two ES~\cite{bauer}, are predicted in the regime of
high imbalance.

Here we apply a quasi-Corbino geometry\cite{alida} to study
electron transport between spin-resolved ES at high imbalance. We
find a pronounced hysteresis in the $I-V$ curves, originating from
slow relaxation processes. We determine characteristic relaxation
times to be 25 s and 200 s which points to the presence of two
different relaxation mechanisms. These results are discussed in
terms of the dynamic nuclear polarization of the edge region via
the hyperfine interaction.

\section{Samples and experimental technique}
The samples are fabricated from two molecular beam epitaxial-grown
GaAs/AlGaAs heterostructures with different carrier concentrations
and mobilities. One of them (A) contains a 2DEG located 70~nm
below the surface.  The mobility at 4K is 800 000 cm$^{2}$/Vs and
the carrier density 3.7 $\cdot 10^{11}
 $cm$^{-2}$.
For heterostructure B the  corresponding parameters are 110~nm,
2.2 $\cdot 10^{6}$cm$^{2}$/Vs and 1.35$\cdot 10^{11}$cm$^{-2}$. We
obtain similar results on samples of both materials. For this
reason we restrict the discussion here to results obtained from
samples of wafer A.

\begin{figure}
\includegraphics*[width=\columnwidth-3cm]{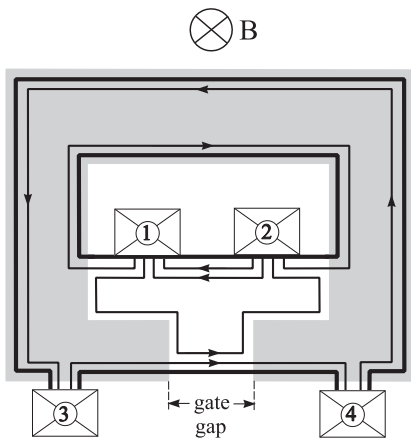}%
\caption{Schematic diagram of the pseudo-Corbino geometry.
Contacts are positioned along the etched edges of the ring-shaped
mesa. The shaded area represents the Schottky-gate. Arrows
indicate the direction of electron drift in the edge channels for
the outlined configuration: filling factors are $\nu=2$ in the
ungated regions and $g=1$ under the gate. \label{fig1}}
\end{figure}

Samples are patterned in a quasi-Corbino geometry~\cite{alida}
(see Fig.~\ref{fig1}). The square-shaped mesa has a rectangular
etched region inside. Ohmic contacts are made to the inner and
outer edges of the mesa. The top gate does not completely encircle
the inner etched region but leaves uncovered a narrow ($3\mu$m)
strip (gate-gap) of 2DEG at the outer edge of the sample.

In a quantizing magnetic field at integer filling factors (e.g.
$\nu=2$, see Fig.~\ref{fig1}) edge channels are running along the
etched edges of the sample. Depleting the 2DEG under the gate to a
smaller integer filling factor (e.g. $g=1$, as shown in the
figure) some channels ($\nu-g$) are reflected at the gate edge and
redirected to the outer edge of the sample. In the gate-gap
region, edge channels originating from different contacts run in
parallel along the outer (etched) edge of the sample, on a
distance determined by the gate-gap width. Thus, the applied
geometry allows us to separately contact edge states and bring
them into an interaction on a controllable length.  A voltage
applied between inner and outer  ohmic contacts makes it possible
to produce a significant imbalance between edge channels because
the gate-gap width of a few microns is much smaller than the
typical equilibration length between ES (more than $100\mu$m
 at low temperatures~\cite{muller,haug,alida}).

\begin{figure}
\includegraphics*[height=\textheight-9cm]{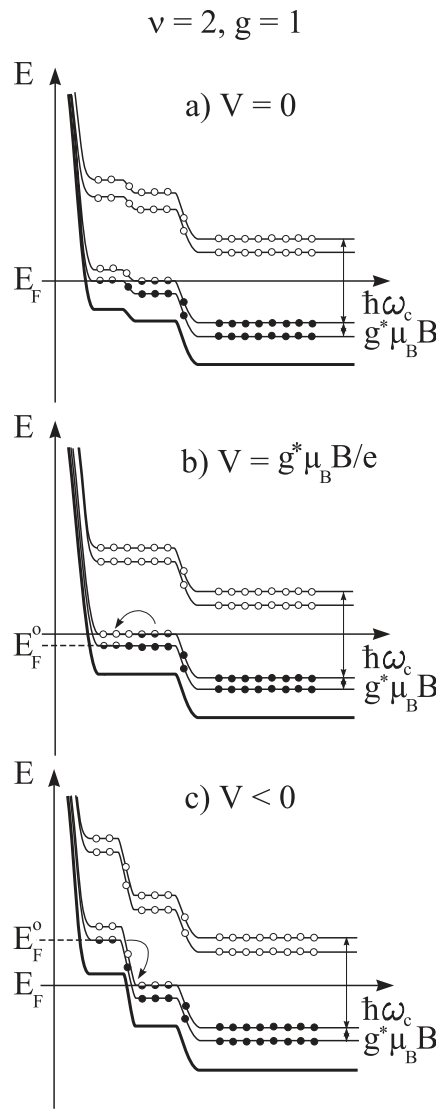}%
\caption{ Energy subband diagram of the sample edge in the
gate-gap. a) No voltage $V$ applied between inner and outer edge
states. b) $V>0$, in the situation shown, the outer edge state is
shifted down in energy by $eV=-g^*\mu_B B$. The potential profile
is flattened between inner and outer edge states. c) $V<0$, the
energy shift is equal to $eV>0$. \label{fig6}}
\end{figure}

In our experimental set-up, a positive bias $V>0$ moves the outer
ES down in energy with respect to the inner one (see
Fig.~\ref{fig6} b)) (one inner ohmic contact is grounded).
Therefore, a small positive bias flattens the edge potential
profile between outer and inner edge states. At voltages close to
the energy which separates the involved edge states,  the
potential barrier between edge states disappears and a significant
current starts to flow~\cite{dixon,alida}. In contrast, a negative
bias steepens the potential relief (see Fig.~\ref{fig6} c)), so
that electrons at any negative bias have to tunnel through the
magnetic-field induced barrier. Experimental $I-V$ traces are
expected to be nonlinear and asymmetric with a characteristic
onset voltage on the positive branch, roughly equal to the
corresponding energy gap (for a more thorough discussion
see~\cite{alida}). A current in this case directly reflects a
charge transfer between edge channels in the gate-gap in contrast
to the conventional Hall-bar geometry with crossing
gates~\cite{muller}. Whereas, no clear onset behaviour can be seen
for negative applied voltages.

Adjusting both, magnetic field $B$ and gate voltage $V_g$, it is
possible to change the number of ES in the gate-gap region (equal
to the bulk filling factor $\nu$) and - independently - the number
$g$ of ES transmitted under the gate. Thus the applied geometry
allows us to study transport between spin-split or cyclotron-split
edge channels depending on the adjusted filling factors $\nu$ and
$g$.

We  obtain $I-V$ curves from {\em dc} four-point measurements at a
temperature of 30 mK in magnetic fields up to 16 T. The measured
voltages $V$ are always much smaller than the gate voltage, so the
electron density under the gate is unchanged during the $I-V$
sweeps. The results presented in the paper are temperature
independent below~0.5~K.

\section{Experimental results}

\begin{figure}
\includegraphics[width=\columnwidth]{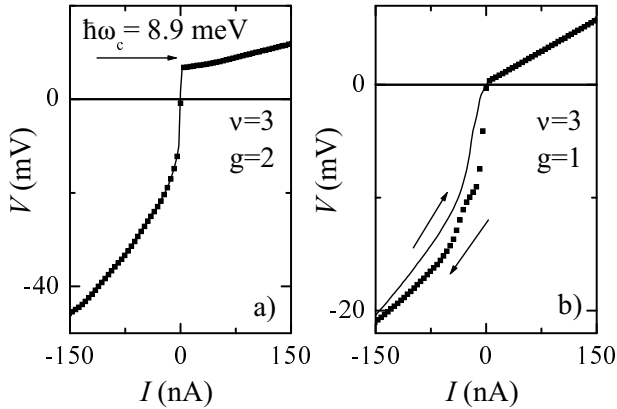}%
\caption{  $I-V$ curves for filling factor combinations a) $\nu=3,
g=2$ (cyclotron splitting)  and b) $\nu=3, g=1$ (spin splitting).
The solid line indicates a sweep from negative to positive
currents and dots represent the reverse sweep direction. The arrow
in a) denotes the theoretical value of the cyclotron gap, whereas
the arrows in b) exhibit the sweep directions. For the dotted
curves the number of points is reduced by 10 times for reasons of
clarity. The magnetic field is 5.2~T. \label{fig2}}
\end{figure}

A typical $I-V$ curve is shown in Fig~\ref{fig2} a) for the
filling factor combination $\nu=3, g=2$. The $I-V$ trace reflects
transport in the gate-gap between cyclotron-split edge states. It
is strongly non-linear and asymmetric with a positive onset
voltage close to the value of the cyclotron energy~\cite{alida}.
The negative branch of the trace changes its slope at a voltage
also comparable to $\hbar\omega_c/e$, due to the crossing of the
outer (partially filled) ES with the excited (empty) level in the
inner one.

Figure~\ref{fig2} b) shows an $I-V$ curve for the filling factor
combination $\nu=3, g=1$ which corresponds to transport between
spin-split ES. The onset voltage on the positive branch is much
smaller in this case, because of the smaller value  of the spin
gap in comparison to the cyclotron gap. However, the most
important difference from Fig.~\ref{fig2} a) is a large hysteresis
for the negative branch of the $I-V$ curve.

The curves in Fig.~\ref{fig2}  are obtained  by continuous sweep
from positive to negative currents and vice versa. Increasing the
sweep rate increases the hysteresis effect. This indicates that
the hysteresis is the result of a
 long-time relaxation process with a characteristic time comparable
  to the sweep time of about ten minutes.

  The described behaviour is contrary to the one observed for transport
  through the cyclotron splitting (see Fig.~\ref{fig2} a))
where there is no hysteresis effect discernible. Because of the
much smaller bulk 2DEG dissipative conductivity under the gate for
cyclotron split filling factors, the obtained hysteresis can not
be caused by the charging of the bulk 2DEG. This fact was checked
for different filling factor combinations on samples from two
different wafers: the hysteresis is present only for transport
between spin-split edge channels and there is no hysteresis in the
$I-V$ curves corresponding to cyclotron splitting.

\begin{figure}
\includegraphics[width=\columnwidth]{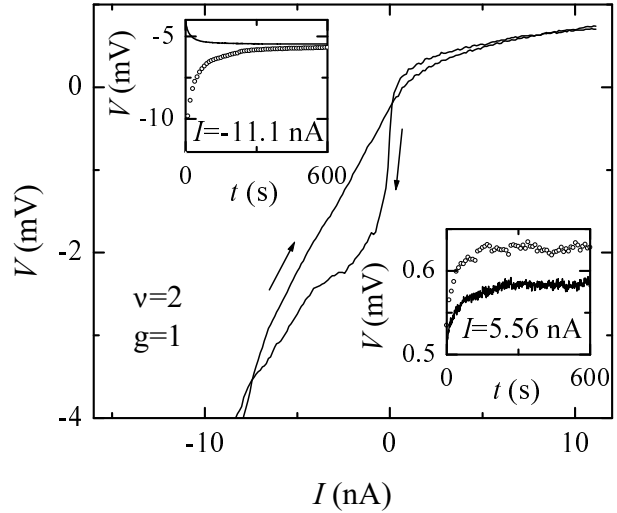}%
\caption{  $I-V$ curves for the filling factor combination $\nu=2,
g=1$ (spin splitting) for small biases. The two different sweep
directions are indicated by arrows.  Insets show the relaxation
curves at fixed currents $I=-11.1$~nA (left inset) and $I=5.56$~nA
(right inset) obtained for two dwelling currents
$I^-_{dwell}=-222$~nA (solid curves) and $I^+_{dwell}=111$~nA
(dotted ones). The magnetic field is 7.7~T. \label{fig3}}
\end{figure}

The zero-bias region of the $I-V$ trace is shown in
Fig.~\ref{fig3} for the filling factor combination $\nu=2, g=1$. A
small hysteresis can also be observed for the positive branch. The
asymmetry of the $I-V$ curve is also reflected in the hysteresis:
at negative biases it is more pronounced, while for the positive
branch there is only a small hysteresis observable at very small
currents.

To directly investigate the time dependence of the relaxation, we
measure the change of the voltage drop at different fixed
currents. To prepare a stable state of the system, a dwelling
current $I_{dwell}$ is applied for a time long enough (about 10
minutes) to observe a stable voltage drop. This procedure provides
a reproducible initial state of the system. Directly switching to
a current $I$ after the dwell, we measure the time-dependent
voltage
 drop $V(t)$. The resulting $V(t)$ curves are well reproducible.

 Examples of these $V(t)$ dependencies are shown in the
insets of Fig.~\ref{fig3}. For both, positive and negative
branches of the $I-V$ trace, time-dependent relaxation is measured
after dwelling at two different currents $I^{+}_{dwell}=111$~nA
(circles) and $I^{-}_{dwell}=-222$~nA (solid curves). It can be
seen that both branches of the $I-V$ traces differ not only in the
size of the relaxation (by two orders of magnitude), but also in
the dependence on the sign of the dwelling current. For the
positive branch ($I>0$) $V(t)$ curves are qualitatively
independent of the sign of $I_{dwell}$ and the relaxation always
appears as a rising of the resistance. For the negative branch
($I<0$), on the other hand, the resistance is increasing with time
for negative $I^{-}_{dwell}<0$ and decreasing for a positive one
$I^{+}_{dwell}>0$. Thus, for the negative branch of the $I-V$
trace the character of the relaxation as well as it´s starting
value are very sensitive to the sign of $I_{dwell}$.

\begin{figure}
\includegraphics[width=\columnwidth]{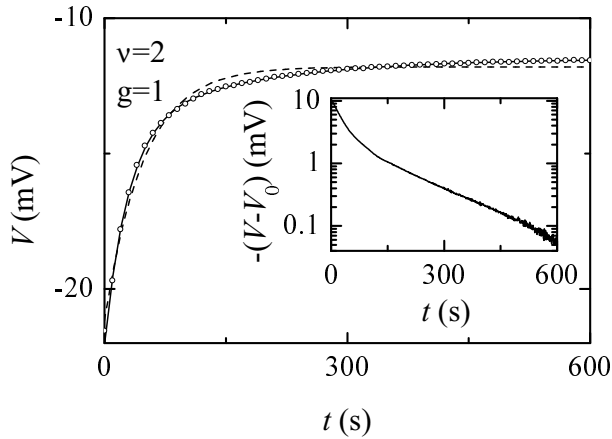}%
\caption{  Relaxation curve for the filling factor combination
$\nu=2, g=1$ at $I=-22.2$~nA for a positive dwelling current
$I^{+}_{dwell}=111$~nA (circles). The number of points is
diminished by 10 times for clarity. The solid curve indicates a
fit by a double exponential decay function
(Eq.~\protect\ref{eq1}), the dashed one is a fit by one
exponential function only. The inset shows the same experimental
relaxation curve in a semilogarithmic plot shifted by
$V_0=V(t=\infty)=-11.5$~mV.\label{fig4}}
\end{figure}

 The experimental $V(t)$ curves seem  to obey an exponential law of
relaxation but clearly consist of two different regions. We find
that the relaxation curves for transport between two spin-split
edge channels at negative currents can well be fitted by a
double-exponential decay

\begin{equation}
V(t)=V_0+V_1exp(-\frac{t}{\tau_1})+V_2exp(-\frac{t}{\tau_2}),
\label{eq1}
\end{equation}

as shown in Fig.~\ref{fig4} (solid curve). For a comparison, a
single-exponential fit (dashed curve in Fig.~\ref{fig4}) is given,
which cannot describe adequately the experimental data, especially
for $t>50$~s. The inset in semilogarithmic axes demonstrates
clearly the presence of the second exponential dependence which
especially dominates the long-time behavior.

The decay times obtained from the double-exponential fit are of
the order of $\tau_1 \sim 25s$ and $\tau_2 \sim 200s$. They are
practically independent of the dwelling current and the
measurement current. Measurements of the relaxation in tilted
fields show that the time constants $\tau_1$ and $\tau_2$ are also
independent of the in-plane magnetic field.  For positive currents
(positive branch of the $I-V$ trace) the accuracy of the
determination of $\tau_2$ is smaller than for negative ones
because of the smaller value of the relaxation (see inset to
Fig.~\ref{fig3}).

\begin{figure}
\includegraphics[width=\columnwidth]{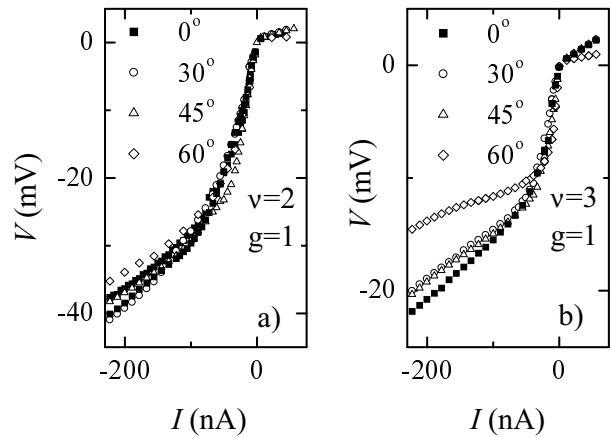}%
\caption{ Steady-state $I-V$ curves for normal and tilted magnetic
fields for a) $\nu=2, g=1$ and b) $\nu=3, g=1$. The tilt angles
shown are $0^\circ$, $30^\circ$, $45^\circ$, and $60^\circ$. For
the filling factor combination $\nu=2, g=1$ at normal magnetic
field the results of two different cooling cycles of the sample
are presented. \label{fig5}}
\end{figure}

From the $V(t)$ curves the time-independent, steady state of the
system can be extrapolated. It can be obtained either from $V_0$
as a fitting parameter (see Eq.~(\ref{eq1})) or as the last value
of the relaxation curve at $t=600$~s.  The difference is
negligible. The resulting steady state $I-V$ traces are presented
in Fig.~\ref{fig5} for two different spin-split filling factor
combinations for normal and tilted magnetic fields. It can be seen
that the stationary $I-V$ curves for $\nu=2, g=1$ (Fig.~\ref{fig5}
a)) are independent of the in-plane-component of the magnetic
field. The deviations are of the same order as  the difference
between two different cooling cycles at zero tilt angle.

 For other spin-split filling factor combinations (such as $\nu=3, g=1$) a
larger number of edge channels is involved in the transport.
Besides charge transfer among spin-split ES in the gate-gap,
transport between cyclotron-split edge channels has to be taken
into account. The cyclotron-splitting is well
known~\cite{cycltilt} to be dependent on the in-plane magnetic
field, which is also verified for this sample.

Therefore, there is some influence of the in-plane magnetic field
observable on the steady-state $I-V$ traces (see Fig.~\ref{fig5}
b)) for these filling factor combinations. The resistance is
decreasing
 with increasing in-plane magnetic field because of the reduction
 of the cyclotron gap.

\section{Discussion}

For electron transfer between spin-resolved ES it is necessary to
change both the spin and the spatial position of the electron. For
this reason, we consider three possible mechanisms  for the
electron transfer: (i) magnetic impurities, (ii) spin-orbit
interaction, and (iii) hyperfine interaction. An influence of the
magnetic impurities can be excluded because of the high quality of
the MBE process for GaAs/AlGaAs and taking into account the fact
that our samples, which  are grown in two different MBE systems,
exhibit similar behavior.

The spin-orbit interaction is well known to be responsible for
electron transfer between ES at small
imbalance~\cite{muller,khaetskii}.It should also be taken into
account for our samples, that some part of the electrons will
relax in this way. However, spin-orbit coupling cannot explain a
voltage relaxation on the macroscopic time scale of the order of
$\tau_1 \sim 25 s$. The observed voltage relaxation is also
independent of the in-plane magnetic field component, as both the
relaxation times $\tau_1,\tau_2$ and the steady-state of the
system are insensitive to the in-plane magnetic field. This is
very unusual for the spin-orbit coupling and indicates a different
origin of the relaxation for the transport between spin-resolved
edge channels. On the other hand, the obtained relaxation time
$\tau_1\sim 25$~s is close to the nuclear spin relaxation times in
GaAs (of the order of 30~s~\cite{dixon,komiyama}). For this
reason, the relaxation can be attributed to the hyperfine
interaction.

It is a well known fact that the hyperfine interaction in
GaAs/AlGaAs heterostructures is strong enough to have an influence
on transport experiments~\cite{dobers,berg,portal}. A Hamiltonian
of the hyperfine interaction can be written as
\begin{equation}
A\bm{I \cdot S}=\frac{1}{2}(I^+S^- + I^-S^+)+ A S_z I_z,
\label{eq2}
\end{equation}
where $A>0$ is the hyperfine constant, $\bm{I}$ is the nuclear
spin and $\bm{S}$ is the electron spin.

 At the temperature of
the experiment (30 mK) in a magnetic field (below 16 T) the static
nuclear polarization $\langle I_z \rangle$ by the external
magnetic field is negligible. Nevertheless, a significant  {\em
dynamic} polarization of the nuclei is possible: an electron
spin-flip causes the spin-flop of a nucleus in the GaAs lattice as
described by the first term in Eq.~\ref{eq2} (the so called
flip-flop process). Thus, a current flow between spin-resolved
edge channels produces a dynamic nuclear polarization (DNP)
$\langle I_z \rangle$  in the edge region of the sample
~\cite{dixon,komiyama}. This polarization affects the electron
energy through the second term in the Eq.~\ref{eq2} (the
Overhauser shift). The influence of the nuclear polarization on
the electron energy can be conveniently described by the effective
Overhauser field $B_{Ov}=A\langle I_z \rangle/g^{*}\mu_B$ which
affects the Zeeman splitting $g^{*}\mu_B(B+B_{Ov})$.

A different relaxation behavior for positive and negative currents
in our experiment can be qualitatively  understood in terms of
DNP. Let us discuss the filling factor combination $\nu=2, g=1$
(Fig.~\ref{fig6}). Because of the negative effective $g$-factor
($g^{*}=-0.44$ in bulk GaAs), electron spins in the outer ES are
polarized in the field direction ("up"-polarization) while in the
inner ES they are polarized "down".

 A negative applied
bias shifts the outer ES up in energy with respect to the inner
one (see Fig.~\ref{fig6} c)). Electrons tunnel through the
incompressible strip between outer and inner ES with a spin-flip
from up to down. Some of these electrons relax due to the
spin-orbit coupling, changing their energy by phonon emission.
Nevertheless, inside the incompressible strip horizontal (in
energy) transitions from the filled to empty states are possible
due to the flip-flop process. The electron spin-flip from up to
down leads to a nuclear spin-flop from down to up. Thus, a current
persisting for a long time induces a DNP  $\langle I_z \rangle >
0$ in the gate-gap. Because of the negative $g^{*}$-factor in
GaAs, the effective Overhauser field is antiparallel to the
external field $B_{Ov} < 0$ and decreases the value of the Zeeman
splitting $g^{*}\mu_B(B+B_{Ov})$.

For a positive bias exceeding the onset voltage $V_{on} \sim
g^{*}\mu B$, there is no more potential barrier for electrons
between edge states (see Fig.~\ref{fig6} b) ). Electrons are
flowing from the inner state to the outer ES and rotate the spin
in vertical transitions afterwards (e.g. by emitting a photon),
possibly far from the gate-gap region. Nevertheless, for a bias $V
> 2V_{on}$ electrons can also tunnel from the filled (spin up)
state to the empty one (spin down) in the incompressible strip due
to the flip-flop mechanism, relaxing later to the ground state
vertically. This flip-flop also produces an "up" nuclear
polarization $\langle I_z \rangle
> 0$ in the gate-gap accompanied by an Overhauser field $B_{Ov} < 0$
antiparallel to the external magnetic field, which diminishes the
Zeeman splitting.

The value of the net nuclear polarization $\langle I_z \rangle$ is
therefore determined by the current connected with flip-flop
processes which in turn is controlled by the applied bias $V$ (see
Fig.~\ref{fig6} c) ). Thus, after dwelling at a positive current
and switching to a negative one the nuclear polarization (i.e. the
Overhauser field) is increasing significantly during the
relaxation process, diminishing the Zeeman splitting. As the
Zeeman splitting determines the spatial distance between
spin-split ES, the tunneling length for the electrons decreases
during the relaxation process~\cite{komiyama}. For this reason, in
the described situation, the relaxation goes along with a decrease
of the resistance, as seen in the experiment (left inset to
Fig.~\ref{fig3} (dots)).

Dwelling  at a high negative current and switching to a lower one
(i.e. closer to the zero), the nuclear polarization is diminishing
in value. The corresponding change in the Overhauser field
increases the Zeeman splitting. As a result, the tunneling length
is rising, leading to an increase of the resistance, as depicted
by the solid curve in the left inset to Fig.~\ref{fig3}.

 After a dwell at a
negative or large positive current and switching to a small
positive one, the nuclear polarization is always diminishing in
value, leading to an increase of the resistance, as can be seen in
the right inset to Fig.~\ref{fig3}.

The proposed picture is thus in qualitative agreement with the
experimental data. The relaxation should take place on a time
scale determined by the nuclear spin-lattice relaxation time $T_1$
and obey an exponential law.

The presence of the two different relaxation times in the
experimental traces can result directly from the injection of
spin-polarized electrons in the gate-gap region of the applied
experimental geometry (Fig.~\ref{fig6}, for a model calculation of
a similar problem see~\cite{pershin}). The characteristic time
scale for establishing the Overhauser field in the gate-gap is
governed by the applied current and the diffusion of the nuclear
polarization from the gate-gap region because of nuclear spin-spin
interactions. A combination of these two processes should be
responsible for the first relaxation with the characteristic time
$\tau_1 \sim 25$~s. However, diffusion takes place on length
scales much larger than the gate-gap width but smaller than the
sample size. The second relaxation process with the characteristic
time $\tau_2$ of the order of the nuclear spin-lattice relaxation
time $T_1$  is therefore attributed to the establishing of a
stable nuclear polarization outside the gate-gap.

For measurements at constant current the relaxation in voltage
reflects not only the change of the spatial positions of
spin-split ES, but also directly the Overhauser shift. The latter
is negligible for the  strong relaxation at negative biases, but
becomes significant for positive ones. From the  value of the
relaxation in this case we can estimate the Overhauser shift
$A\langle I_z\rangle\langle S_z\rangle~\sim\Delta V^{+} \sim
100\mu$~V, which gives a value of the Overhauser field
$B_{Ov}=\Delta V^{+} / (g^* \mu_B) \sim 4$~T. This value is
smaller than the highest Overhauser field (5.3~T) reported for
GaAs~\cite{safarov} and close to the value reported for DNP in
heterostructures~\cite{dixon}. On the other hand, $B_{Ov}=4$~T is
strong enough to significantly change the Zeeman splitting in the
external field of 7.7~T (which corresponds to $\nu=2$ in the bulk)
which gives further support for the proposed picture.

\section{Conclusion}
Performing  direct measurements of the electron transport between
spin-split edge states at high imbalance, we found a long-time
relaxation in contrast to charge transfer between cyclotron-split
edge states. The determined characteristic times are of the order
of 25 s and 200 s which points to the presence of two different
relaxation mechanisms. We attribute this relaxation to the
formation of a dynamic nuclear polarization (DNP) near the sample
edge. The presence of the two relaxation mechanisms is interpreted
as the formation of a DNP inside the gate-gap region due to
flip-flop processes and outside it as a consequence of the
diffusion of nuclear spins. We also found that an in-plane
magnetic field has no influence both on the relaxation between two
spin-split edge channels and on the steady-state of the system.

\begin{acknowledgments}
We wish to thank Dr. A.A.~Shashkin for help during the experiments
and discussions. We gratefully acknowledge financial support by
the Deutsche Forschungsgemeinschaft, SPP "Quantum Hall Systems",
under grant LO 705/1-1.  The part of the work performed in Russia
was supported by RFBR,  the programs "Nanostructures" and
"Mesoscopics" from the Russian Ministry of Sciences. V.T.D.
acknowledges support by A. von Humboldt foundation.
\end{acknowledgments}

\end{document}